\documentclass[amsmath,latexsym]{article}
\usepackage{latexsym}
\usepackage{graphicx}
\begin{document}
\begin{center}{\bf CASIMIR FORCE ON A MICROMETER SPHERE IN A DIP:\\
PROPOSAL OF AN EXPERIMENT}

\vspace{1cm} I. Brevik \footnote{E-mail: iver.h.brevik@ntnu.no},
E. K. Dahl\footnote{E-mail: eskil.dahl@phys.ntnu.no}, and G. O.
Myhr\footnote{E-mail:  myhr@fysmat.ntnu.no}

\bigskip
\bigskip

Department of Energy and Process Engineering, Norwegian University
of Science and Technology, N-7491 Trondheim, Norway

\begin{abstract}
The attractive Casimir force acting on a micrometer-sphere
suspended in a spherical dip, close to the wall, is discussed.
This setup is in principle directly accessible to experiment. The
sphere and the substrate are assumed to be made of the same
perfectly conducting material.

\end{abstract}
December 2004

\end{center}


\section{Introduction}

Experiments aiming at testing the theory of the Casimir effect
(\cite{casimir48}; for recent reviews see
\cite{milton04,milton01,bordag01}) are more numerous than what one
might perhaps think. Let us here only highlight some examples,
starting with the classic experiment of Sparnaay
\cite{sparnaay58,sparnaay89}. This experiment tested the Casimir
force between two parallel plates, made of chromium steel,
chromium, and aluminium. With the exception of aluminium (whose
problems most likely were due to impurities), the results were in
good qualitative agreement with the Lifshitz formula
\cite{lifshitz56}, calculated from the assumption of perfect
reflecting boundaries. The experimental technique was based upon
use of a spring balance (sensitivity about $10^{-3}$ dynes),
sensing the attractive force. The plates were assigned parallel by
visual inspection.

Another well known classic experiment is the one of Sabisky and
Anderson \cite{sabisky73}, dealing with the properties of liquid
helium films adsorbed on cleaved surfaces of alkaline-earth
fluoride crystals at $T=1.38$ K. Film thicknesses measured by
means of acoustic interferometry, were found to lie between 1 and
25 nm. At thermal equilibrium the film thickness gets a value that
is determined thermodynamically, given the Lifshitz formula for
the Casimir
 force as input. The results measured were in very good
 quantitative agreement with the Lifshitz expression.

 The modern series of experiments was initiated with the seminal
 work of Lamoreaux \cite{lamoreaux97}. He used a balance based on a
 torsion pendulum to measure the Casimir force between a gold
 coated spherical lens (radius about 12 cm) and a flat plate. The
 lens was mounted on a piezo stack and the plate on one arm of the
 torsion balance. The Casimir force would result in a torque,
 which was detected via a capacitance measurement. Maximum
 separation between the two surfaces was 12.3 $\mu$m. In a later
 note, Lamoreaux included corrections, such as those arising from
 finite conductivity \cite{lamoreaux98}, and with Buttler \cite{lamoreaux04}
 he gave recently an analysis of thermal noise in torsion pendulums.
  The Lamoreaux experiment
 gave rise to a surge of activity, both experimentally and
 theoretically.

 In the most recent years, the atomic force microscope (AFM) used in
 particular by Mohideen et al. \cite{mohideen98,roy99,harris00}
 has led to the most accurate determination of the Casimir force
 between a micrometer-sized sphere and a plate. By using a
 sphere/plate configuration, one avoids the strict requirement
 about parallelism that is so demanding in the case of parallel
 plates. The accuracy is now of the order of a few per cent; this
 accuracy being actually under current debate mainly because of
 the temperature corrections.

 We shall not here go into further detail as regards the
 experimental status. A  detailed exposition on the experiments up to 2001 is given
 in the review of  Bordag et al. referred to earlier \cite{bordag01}, and a detailed survey of
 the developments in the last four years is given in Milton's review \cite{milton04}, Sect. 3.6.
  We mention, though, the
 impressive plane-plate experiment of Bressi et al.
 \cite{bressi02}; they were able to guarantee a parallelism of the plates to better
 than $3\times 10^{-5}$ rad. (It may even be that this experiment has been the first  to
 measure the temperature corrections to the Casimir force; cf. the
 discussions on temperature corrections in \cite{hoye03,brevik04}.)

 The main purpose of the present note is to  propose a new variant of the
 sphere-substrate configuration, namely a metal sphere suspended
 in a spherically formed metallic dip. See figure 1. We will make
 a simple, approximate, calculation of the vertical Casimir force
 on the sphere, utilizing the known theory for the Casimir effect
 under conditions of perfect spherical symmetry. There exist
 several theoretical treatments of the Casimir effect under
 conditions of spherical symmetry - cf.
 \cite{milton78,brevik94,hoye01,brevik02,brevik02a} for instance - but the
 experimental tests of this kind of Casimir forces have so far
 been absent. Leaving aside practical difficulties such as the need of keeping the
  sphere in the dip in a stable lateral position, we hope nevertheless that the
  present simple idea can be of interest to experimentalists.

  \begin{figure}[htbp]
    \centering
    \resizebox{0.6\textwidth}{!}{\includegraphics{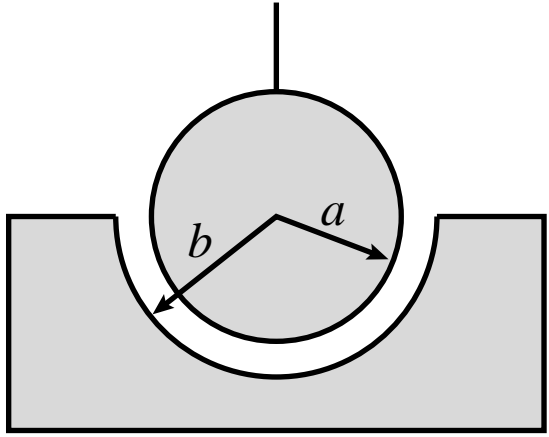}}
    \caption{The proposed experimental setup. A metal sphere of radius
      $a$ is suspended in a spherically formed metallic dip of radius $b$.}
    \label{fig:setup}
  \end{figure}

  Temperature corrections are  not included in the main formalism but are discussed
  in section 3.  We use natural
  units, $\hbar =c=1$, in the intermediate calculations, and we
  employ Heaviside-Lorentz units.

\section{Geometrical Set-Up}

We begin by assuming perfect spherical symmetry: let there be two
concentric perfectly conducting singular shells situated at $r=a$
and $r=b$. We shall be interested in the distribution of fields in
the annular region $a<r<b$, at $T=0$. The most natural way of
approach when describing this situation is to make use of the
Green functions; for the case of spherical symmetry this kind of
theory was worked out by Milton {\it et al.} \cite{milton78}.
There occur two scalar Green functions in the problem, $F_l(r,r')$
and $G_l(r,r')$. As shown in \cite{brevik94} for the double-shell
situation, the electromagnetic boundary conditions at $r=a,b$
transform into the following conditions for the scalar Green
functions:
\begin{equation}
F_l(a,r')=0,\quad \frac{{\rm d}}{{\rm d}r}\left[rG_l(r,r')\right]_{r=a}=0,
\label{1}
\end{equation}
and similarly for $r=b$.

The surface force density, here called $f_b$, on the outer surface
$r=b$ is calculated from Maxwell's stress tensor. To this end we
need the two-point functions, which are in turn given by the
components of the Green functions. Some calculation yields for the
radial two-point function for the electric field
\begin{equation}
\langle E_r^2(b-)\rangle=\frac{1}{\pi b^4}\int_0^\infty \frac{{\rm
d}y}{y} \sum_{l=1}^\infty \frac{2l+1}{4\pi}l(l+1)
\frac{s_l(y)-A_G(ay/b)e_l(y)}{s_l'(y)-A_G(ay/b)e_l'(y)}, \label{2}
\end{equation}
whereas the corresponding transverse function for the magnetic
field becomes
\begin{equation}
\langle H_\perp (b-)\rangle =-\frac{1}{\pi b^4}\int_0^\infty y{\rm
d}y \sum_{l=1}^\infty \frac{2l+1}{4\pi}\left[
\frac{s_l(y)-A_G(ay/b)e_l(y)}{s_l'(y)-A_G(ay/b)e_l'(y)}+
\frac{s_l'(y)-A_F(ay/b)e_l'(y)}{s_l(y)-A_F(ay/b)e_l(y)} \right].
\label{3}
\end{equation}
Here $\langle E_r^2(b-)\rangle$ is shorthand for $\langle
E_r(r)E_r(r')\rangle_{r'\rightarrow r=b-}$, etc.; $y$  means the
nondimensional frequency $y=\hat{\omega}b$  with $\hat{\omega}$
being the Wick-rotated frequency, and $s_l(x)=\sqrt{\pi
x/2}\;I_\nu (x)$, $e_l(x)=\sqrt{2x/\pi}\;K_\nu$ with $\nu=l+1/2$
are the Riccati-Bessel functions defined such that their Wronskian
is $W\{s_l,e_l\}=-1$. Prime means derivative with respect to the
whole argument. The coefficients $A_F$ and $A_G$ in Eq.~(\ref{3})
are
\begin{equation}
A_F(x)=\frac{s_l(x)}{e_l(x)},\quad A_G(x)=\frac{s_l'(x)}{e_l'(x)}.
\label{4}
\end{equation}
The reason why the integration over $y$ runs to infinity in
Eq.~(\ref{3}) is that the medium in the surfaces is assumed to be
perfectly conducting (i.e., with permittivity $\varepsilon
\rightarrow \infty$) for all frequencies.

The other two-point functions are found to vanish,
\begin{equation}
\langle E_\perp^2(b-)\rangle=\langle H_r^2(b-)\rangle=0, \label{5}
\end{equation}
so it is simple to find the surface force density at $r=b$ via
use of the Maxwell stress tensor:
\[ f_b=\frac{1}{2}\left[
-\langle E_r^2(b-)\rangle+\langle H_\perp^2(b-)\rangle
\right]\bf{\hat{r}}
\]
\begin{equation}
=\frac{-1}{2\pi b^4}\int_0^\infty y{\rm d}y \sum_{l=1}^\infty
\frac{2l+1}{4\pi}\left[ \frac{s_l'(y)-A_F(ay/b)e_l'(y)}
{s_l(y)-A_F(ay/b)e_l(y)}+\frac{s_l''(y)-A_G(ay/b)e_l''(y)}
{s_l'(y)-A_G(ay/b)e_l'(y)}\right] \bf{\hat{r}}. \label{6}
\end{equation}
The surface force $F_z$ that we are interested in, is the $z$
component of the surface force density integrated over the lower
hemisphere, $\frac{1}{2}\pi < \theta < \pi$, $0< \phi <2\pi$. This
integration is trivial, since the magnitude of the surface force
density  contains no angular dependence. It is moreover convenient
to rewrite $F_z$ in such a way that the mutual contribution is
separated out. Some formal manipulations yield
\[ F_z=\frac{1}{2b^2}\int_0^\infty y{\rm d}y\sum_{l=1}^\infty
\frac{2l+1}{4\pi}
\left[\frac{s_l'(y)}{s_l(y)}+\frac{s_l''(y)}{s_l'(y)} \right] \]
\begin{equation}
+\frac{1}{2b^2}\int_0^\infty y{\rm d}y\sum_{l=1}^\infty
\frac{2l+1}{4\pi} \frac{\partial}{\partial y}\ln
\left[\left(1-A_F(x)\frac{e_l(y)}{s_l(y)}\right) \left(
1-A_G(x)\frac{e_l'(y)}{s_l'(y)}\right)\right]. \label{7}
\end{equation}
Here, $x = ay/b$ is a function of $y$, but when taking the partial
derivative with respect to $y$ in the last term, $x$ and $y$ are
regarded as independent variables. The first term describes the
self-force on the surface $r=b$, due to the fluctuating fields in
the annular region. [If we were taking into account the
contribution from the outer region $r>b$ also, as would strictly
speaking be necessary when considering a double spherical shell,
then there would be an analogous term $e_l'/e_l+e_l''/e_l'$ in
addition; cf. Ref.~\cite{milton78}.] In our case, self-forces are
not of interest. The physically important force thus becomes
\begin{equation}
F_z=\frac{1}{2b^2}\int_0^\infty y{\rm d}y\sum_{l=1}^\infty
\frac{2l+1}{4\pi} \frac{\partial}{\partial y}\ln
\left[\left(1-A_F(x)\frac{e_l(y)}{s_l(y)}\right) \left(
1-A_G(x)\frac{e_l'(y)}{s_l'(y)}\right)\right]. \label{8}
\end{equation}
The expression is positive, corresponding to an upward directed
force on the outer wall $r=b$, which in turn means a downward
directed force on the sphere. The expression is finite as it
stands; no regularization procedure is necessary.

To make the expression more practically useful, we employ the
Debye expansion for the Riccati-Bessel functions. The calculation
is parallel to that in Ref.~\cite{brevik94}, and will not be
repeated here. We give the result in dimensional form, when
reintroducing $a$ instead of $b$ and working to the first order in
the small quantity $d/a$,
\begin{equation}
F_z=\frac{\pi^2\hbar c}{240 d^4} (\pi
a^2)\left(1+\frac{4}{3}\frac{d}{a}\right). \label{9}
\end{equation}
Here we have separated out the standard expression $\pi^2\hbar
c/240 \,d^4$ for the Casimir surface force density between flat
parallel plates. It is seen from Eq.~(\ref{9}) that the
curvilinear geometry leads to a slightly increased force as
compared with the force between parallel plates having an
effective area of $\pi a^2$. For instance, if $a=50\; \mu$m,
$d=1\; \mu$m, the last correction term in Eq.~(\ref{9}) amounts to
2.6 \%.

\section{Discussion}

The most characteristic property of the present proposal is that
it suggests how the Casimir formalism worked out for spherically
symmetric geometries can be exposed to an experimental test. As
far as we know, this is the first proposal of such a  kind. Of
course, our calculation above is only approximate. Let us make a
few final remarks:

$\bullet$  The most evident simplification that we have made, is
to assume that the field distribution is the same as in the case
of perfect spherical symmetry, all over the dip. Of course, near
$\theta= \frac{1}{2}\pi $ there are "stray" fields making the
distribution different from the perfectly symmetric case. A
circumstance which however diminishes the influence from the stray
fields, is that the $z$ component becomes only slightly influenced
near $\theta = \frac{1}{2}\pi$. We can estimate the magnitude of
the effect by performing the integral over $\theta$ in the $F_z$
calculation from $\frac{1}{2}\pi +\Delta$ to $\pi$, instead of
from  $\frac{1}{2}\pi$ to $\pi$. The result is that the expression
(\ref{9}) gets multiplied with a correction factor $(1-\sin^2
\Delta)$. For example, taking the error introduced by the stray
fields to be $\Delta = 5^0$, we get a decrease in $F_z$ of about
0.8 \%. It would be quite a difficult task to make an accurate
calculation of the influence from the stray fields; one would have
to solve the complicated field distribution problem around
$\theta=\frac{1}{2}\pi$.

$\bullet$  It is of interest to compare our results with the new
technique proposed by Jaffe and Scardicchio based on optical paths
\cite{jaffe04,scardicchio04,scardicchio04a}. This technique
assumes classical optics; it is most accurate at short wavelengths
and where diffraction is not important. The technique has so far
been applied to the case of scalar fields. The main procedure is
to write the Casimir energy as a trace of the Green function; then
the Green function is replaced by the sum over contributions from
optical paths labelled by the number of reflections from the
conducting surfaces. There are two central quantities in the
analysis: first, there is the length $l_r(x)$ of the closed
geometric optics ray beginning and ending at the point $x$ and
reflecting $r$ times from the surfaces; secondly, there is the
so-called enlargement factor $\Delta_r(x)$ of classical optics
associated with the $r$-reflection path beginning and ending at
$x$.

Consider first the simple sphere-plate configuration. The original
wavefront leaving $x$ is spherical. The first reflection from the
sphere produces a new wavefront, with in general two unequal radii
of curvature. When next incident upon the sphere, the asymmetric
wavefront will be transformed in a complicated manner, not yet
worked out even for the scalar field. If we now consider our
proposed experimental setting where there are {\it two} curved
surfaces, this method appears to be quite complicated. We shall
therefore not try to work out this, but it is of interest
nevertheless to compare our results with those obtained for the
sphere-plate configuration.

Let $f^{\rm{optical}}(d/a)$ be the correction factor for a sphere
and a plate, calculated by the optical method. This factor gives
the ratio between the Casimir force and the force obtained for two
parallel plates separated by a gap $d$. From \cite{scardicchio04}
we have, to the first order in $d/a$,
\begin{equation}
f^{\rm optical}(d/a)=1+0.05\, d/a. \label{10}
\end{equation}
We can compare this with the result calculated from the proximity
force approximation (PFA) \cite{blocki77}.  Actually there is an
ambiguity in the PFA: the basic idea of the method is to apply the
parallel-plate result to infinitesimal bits of the (in general)
curved surfaces and integrate them up. The ambiguity lies in which
surface is chosen for the integration. The physically best choice
turns out to be the {\it plate-based} PFA, according to which
\begin{equation}
f^{\rm plate}_{\rm PFA}(d/a)=1-(1/2)d/a. \label{11}
\end{equation}
It is seen that even the signs in the correction terms in
Eqs.~(\ref{10}) and (\ref{11}) are different. Now,
  the sphere-plate situation for the scalar field
has actually been calculated numerically, to a high accuracy
\cite{gies03}. The numerical results show clearly that the
interaction energy increases with increasing values of $d/a$. From
Fig.~4 in \cite{scardicchio04} it is seen that the optical
approximation works well up to $d/a \approx 0.2$.

Finally let us compare these results with our expression (9) for
the Casimir force. We see that our expression corresponds to the
correction factor
\begin{equation}
f(d/a)=1+(4/3)d/a. \label{12}
\end{equation}
The experimental configuration that we have proposed in this paper
is of course different from the sphere-plate configuration with
scalar fields, but we see that our correction term in the factor
$f(d/a)$ is positive. There is thus a qualitative agreement
between our result and the optical method result for scalar
fields, Eq.~(\ref{10}), as well as with the numerical result in
\cite{gies03}.

$\bullet$  We next consider the correction   coming from finite
temperatures. This point is quite subtle. It is instructive to
start with the case where two parallel plates are separated by a
gap $d$. The general condition for applying $T=0$ theory with
reasonable accuracy is that $Td \ll 1$ (in natural units).
Consider, for example, two gold plates at a gap of $d=1\;\mu$m;
from figure 5 in \cite{hoye03} it follows that the surface
pressure is about 1 mPa at $T=300$ K and about 1.15 mPa at $T=0$,
thus a {\it decrease} of 15 \% at room temperature as compared
with zero temperature. As we mentioned earlier, the parallel plate
experiment of Bressi {\it et al.} \cite{bressi02} may even have
been able to measure this temperature effect for the case of
narrow gaps, $d \leq 0.5\; \mu$m. [In this experiment the plates
were actually coated with chromium rather than with gold; when
$d=0.5\; \mu$m corresponding to $Td=0.065$ at room temperature,
the surface pressure is calculated to be 15.5 mPa whereas the
$T=0$ theory yields 20.8 mPa. The theoretically predicted force
reduction is thus somewhat greater than above, about 25 \%; cf.
also the discussion in \cite{hoye03}.]

No consensus  has so far been reached in the literature as regards
the temperature correction, not even in the simple case of
parallel plates. The essential physical point is whether the
transverse electric (TE) mode contributes to the Casimir effect
for a metal in the limit of zero frequency, corresponding to a
Matsubara integer equal to zero. In our opinion it does {\it not},
as spelled out in detail in Ref.~\cite{hoye03}. The papers of
Sernelius and Bostr\"{o}m are in agreement with this opinion
\cite{sernelius04,bostrom00,bostrom00a,sernelius01,sernelius01a}.
It implies, as mentioned,  that the Casimir force is weaker (by
some percent when $d=1\; \mu$m) at room temperature than at zero
temperature. By contrast, the recent paper of Chen {\it et al.}
\cite{chen04}, which is based upon a reanalysis of the earlier
Atomic Force Microscopy (AFM) experiment reported in
\cite{harris00}, claims the temperature correction to be so small
as to be negligible. In that apparatus a gold-coated polystyrene
sphere, mounted on a cantilever of an AFM, was brought close to a
metallic surface and the deflection of the cantilever was measured
as a function of the distance. As a general remark on this
experiment, in spite of the apparent excellent agreement of the
experiment, one suspects that the accuracy of it has been
overestimated. As discussed by Iannuzzi {\it et al.}
\cite{iannuzzi04}, and by Milton \cite{milton04}, at very short
distances $d$ (in the experiment the shortest distance was equal
to 62 nm), the force at $d=62$ nm differs from the force at $d
=62+ \delta$ by more than 3.5 pN (the experimental uncertainty
claimed by the authors) when $\delta$ is larger than a few
angstroms. This implies that $d$ should have been measured with
atomic precision in order to correspond to the accuracy claimed.
One may compare this with the real error in the experiment: the
value of $d$ was determined with  $\pm 1$ nm accuracy. Moreover,
the reason for the claimed smallness of the temperature correction
in \cite{chen04} is that the analysis is based upon the plasma
dispersion relation, which in turn implies that the zero TE mode
contributes to the Casimir force.

After these introductory remarks on the parallel-plate geometry,we
now turn to the finite temperature version of the expression
(\ref{8}). It is convenient to work in terms of the free energy
$F$, instead of the force $F_z$. The relationship between these
quantities is $F_z=\partial F/\partial b$ (recall that $F_z$ is
with our sign conventions positive, whereas the interaction free
energy $F$ has to be negative). The finite temperature version of
$F$ is given by
\begin{equation}
\beta F=\frac{1}{2}{\sum_{m=0}^\infty }' \sum_{l=1}^\infty \nu \ln
\left[\left(1-A_F(x)\frac{e_l(y)}{s_l(y)}\right) \left(
1-A_G(x)\frac{e_l'(y)}{s_l'(y)}\right)\right], \label{13}
\end{equation}
where $\beta =1/T,\, \nu=l+1/2,\, x=2\pi ma/\beta$ and $y=2\pi
mb/\beta$ being the nondimensional frequencies. The prime on the
$m$ summation means that $m=0$ is to be taken with half weight.

We shall consider some limiting cases of this expression. First
consider the {\it static} case, whereby we mean that the frequency
is zero ($m=0$). This means that we take the analytical limits of
$s_l(x)$ and $e_l(x)$ when $x\rightarrow 0$. The result becomes
(cf. Ref.~\cite{brevik02a}):
\begin{equation}
\beta F(m=0)=\frac{1}{2}\sum_{l=1}^\infty \nu \ln \left[
1-\left(\frac{a}{b}\right)^{2\nu} \right]. \label{14}
\end{equation}
In this formula, the radii $a$ and $b$ are arbitrary. The case
where this formula is most useful, is that of moderate or large
gap widths. For comparison, we mention that at room temperature
the Casimir force between a large ($a=300\,\mu$m) gold sphere and
a copper plate gets its dominant  contribution from the zero
frequency  when $d$ is greater than about $0.7\, \mu$m (cf. Fig.~3
in \cite{brevik04a}). The physical reason is that  for large $d$
it is generally the low frequencies that become most important.

Consider next case of finite temperatures, assuming  the gap to be
narrow, $\xi \equiv d/a \ll 1$. Then, from the uniform asymptotic
(Debye) expansions for the Riccati-Bessel functions we obtain, to
the first order in $\xi$ \cite{hoye01}:
\begin{equation}
\beta F= {\sum_{m=0}^\infty}' \sum_{l=1}^\infty \nu\ln\left(
1-e^{-2\xi \sqrt{\nu^2+m^2t^2}} \right), \label{15}
\end{equation}
where $t=2\pi a/\beta$ is the nondimensional temperature. This
expression puts no restriction on the temperature.

Finally let us consider the case of high temperatures. Then, only
the lowest Matsubara frequency $m=0$ contributes, and we are led
back to the expression (\ref{14}). If we in addition require $\xi$
to be small, we obtain \cite{hoye01}
\begin{equation}
\beta F (m=0)=\frac{1}{8\pi}\int_0^\infty q\,dq\,\ln\left(
1-e^{-2qd} \right)=-\frac{\zeta (3)}{32\pi d^2}. \label{16}
\end{equation}
This is the same  as the $m=0$ expression obtained for the free
energy for parallel plates \cite{brevik04a}. (Recall that when the
PFA is assumed, it is this expression that is used for the force
calculation for a sphere/plate system.)

Numerical calculations show in general that there is for low
temperatures a flat plateau for the free energy extending up to
quite high values of $t$. For instance, when $d/a=0.05$ the $T=0$
theory can be used up to $\log_{10}t \approx 1.3$, or $t=20$
(Fig.~6 in \cite{brevik02a}). Since dimensionally $t=2\pi a k_B
T/\hbar c$, this corresponds to $T \approx 140$ K if $a=50\,\mu$m.

For low and moderate temperatures, as mentioned, the Casimir force
diminishes with increasing $T$. For comparison, when dealing with
the force between a gold sphere and a copper plate, we found in
\cite{brevik04} the $T=300$ K force to be weaker than the $T=0$
force by 3.6 \% in the case of a width $d=0.2\,\mu$m. This is
comparable in magnitude with the experimental results of Decca
{\it et al.} \cite{decca03,decca03a}.

\vspace{.3cm}

 To conclude: It should in principle be possible to measure the
expression (\ref{9}) for $F_z$, given that experimental lateral
stability problems for the sphere can be overcome. One should thus
expect to be able to check the predicted effective area factor of
$\pi a^2$. A practical problem is however that the last correction
factor in Eq.~(\ref{9}) appears to be of the same order of
magnitude as the temperature correction.


\newpage

\begin{thebibliography}{99}

\bibitem{casimir48}
Casimir, H. B. G. 1948. {\it Proc. Kon. Ned. Akad. Wetensch.} {\bf
51}, 793.
\bibitem{milton04}
Milton, K. A. 2004. {\it J. Phys. A: Math. Gen.} {\bf 37}, R209.
\bibitem{milton01}
Milton, K. A. 2001. {\it The Casimir Effect: Physical
Manifestations of Zero-Point Energy} (Singapore: World
Scientific).
\bibitem{bordag01}
Bordag, M., Mohideen, U. and Mostepanenko, V. M. 2001. {\it Phys.
Rep.} {\bf 353}, 1.
\bibitem{sparnaay58}
Sparnaay, M. J. 1958. {\it Physica} {\bf 24}, 751.
\bibitem{sparnaay89}
Sparnaay, M. J. 1989. In A. Sarlemijn and M. J. Sparnaay, editors,
{\it Physics in the Making:  Essays on Developments in 20th
Century Physics in Honour of H. B. G. Casimir on the Occasion of
his 80th Birthday} (Amsterdam: North-Holland), p. 235.
\bibitem{lifshitz56}
Lifshitz, E. M. 1956. {\it Zh. Eksp. Teor. Fiz.} {\bf 29}, 94
(Engl. Transl. 1956 {\it Sov. Phys. - JETP} {\bf 2}, 73).
\bibitem{sabisky73}
Sabisky, E. S. and Anderson, C. H. 1973. {\it Phys. Rev. A} {\bf
7}, 790.
\bibitem{lamoreaux97}
Lamoreaux, S. K. 1997. {\it Phys. Rev. Lett.} {\bf 78}, 5.
\bibitem{lamoreaux98}
Lamoreaux, S. K. 1998. {\it Phys. Rev. Lett.} {\bf 81}, 4549(E).
\bibitem{lamoreaux04}
Lamoreaux, S. K. and Buttler, W. T. 2004. arXiv:quant-ph/0408027.
\bibitem{mohideen98}
Mohideen, U. and Roy, A. 1998. {\it Phys. Rev. Lett.} {\bf 81},
4549.
\bibitem{roy99}
Roy, A., Lin, C.-Y.  and Mohideen, U. 1999. {\it Phys. Rev. D}
{\bf 60}, 111101(R).
\bibitem{harris00}
Harris, B. W., Chen, F. and Mohideen, U. 2000. {\it Phys. Rev. A}
{\bf 62}, 052109.
\bibitem{bressi02}
Bressi, G., Carugno, G., Onofrio, R. and Ruoso, G. 2002. {\it
Phys. Rev. Lett.} {\bf 88}, 041804 (2002).
\bibitem{hoye03}
H{\o}ye, J. S., Brevik, I., Aarseth, J. B. and Milton, K. A. 2003.
{\it Phys. Rev. E} {\bf 67}, 056116.
\bibitem{brevik04}
Brevik, I., Aarseth, J. B., H{\o}ye, J. S. and Milton, K. A. 2004.
In K. A. Milton, editor, {\it Proceedings of the 6th Workshop on
Quantum Field Theory Under the Influence of External Conditions},
Paramus, NJ. Rinton Press, p. 54  [arXiv:quant-ph/0311094].
\bibitem{milton78}
Milton, K. A., DeRaad, L. L. Jr., and  Schwinger, J. 1978. {\it
Ann. Phys. (N.Y.)} {\bf 115}, 388.
\bibitem{brevik94}
Brevik, I., Skurdal, H. and Sollie, R. 1994. {\it J. Phys. A:
Math. Gen.} {\bf 27}, 6853.
\bibitem{hoye01}
H{\o}ye, J. S., Brevik, I. and Aarseth, J. B. 2001. {\it Phys.
Rev. E} {\bf 63}, 051101.
\bibitem{brevik02}
Brevik, I., Aarseth, J. B. and H{\o}ye, J. S. 2002. {\it Int. J.
Mod. Phys. A} {\bf 17}, 776.
\bibitem{brevik02a}
Brevik, I., Aarseth, J. B. and H{\o}ye, J. S. 2002. {\it Phys.
Rev. E} {\bf 66}, 026119.
\bibitem{jaffe04}
Jaffe, R. L. and Scardicchio, A. 2004. {\it Phys. Rev. Lett.} {\bf
92}, 070402.
\bibitem{scardicchio04}
 Scardicchio, A. and Jaffe, R. L.
2004. arXiv:quant-ph/0406041.
\bibitem{scardicchio04a}
Scardicchio, A. 2004. arXiv:hep-th/0408013.
\bibitem{gies03}
Gies, H., Langfeld, K. and Moyaerts, L.  2003. {\it JHEP} {\bf
0306}, 018.
\bibitem{blocki77}
Blocki, J., Randrup, J., Swialecki, W. J. and Tsang, C. F. 1977.
{\it Ann. Phys. (N.Y.)} {\bf 105}, 427.
\bibitem{sernelius04}
 Sernelius, Bo E. and Bostr\"{o}m, M.
2004. In K. A. Milton, editor, {\it Proceedings of the 6th
Workshop on Quantum Field Theory Under the Influence of External
Conditions}, Paramus, NJ. Rinton Press, p. 82.
\bibitem{bostrom00}
Bostr\"{o}m, M. and Sernelius, Bo E. 2000. {\it Phys. Rev. A} {\bf
61}, 052703.
\bibitem{bostrom00a}
Bostr\"{o}m, M. and Sernelius, Bo E. 2000. {\it Phys. Rev. Lett.}
{\bf 84}, 4757.
\bibitem{sernelius01}
Sernelius, Bo E. 2001. {\it Phys. Rev. Lett.} {\bf 87}, 139102.
\bibitem{sernelius01a}
Sernelius, Bo E. and Bostr\"{o}m, M. 2001. {\it Phys. Rev. Lett.}
{\bf 87}, 259101.
\bibitem{chen04}
Chen, F., Klimchitskaya, G. L., Mohideen, U., and Mostepanenko, V.
M. 2004. {\it Phys. Rev. A} {\bf 69}, 022117.
\bibitem{iannuzzi04}
Iannuzzi, D., Gelfand, I., Lisanti, M., and Capasso, F. 2004. In
K. A. Milton, editor, {\it Proceedings of the 6th Workshop on
Quantum Field Theory Under the Influence of External Conditions},
Paramus, NJ. Rinton Press, p. 11 [arXiv:quant-ph/0312043].
\bibitem{brevik04a}
Brevik, I., Aarseth, J. B., H{\o}ye, J. S., and Milton, K. A.
2004. arXiv:quant-ph/0410231.
\bibitem{decca03}
 Decca, R. S.,  L\'{o}pez, D. L.,  Fischbach, E., and  Krause, D. E. 2003. {\it Phys.
Rev. Lett.} {\bf 91}, 050402.
\bibitem{decca03a}
Decca, R. S., Fischbach, E.,  Klimchitskaya, G. L.,  Krause, D.
E., L\'{o}pez, D., and  Mostepanenko, V. M. 2003. {\it Phys. Rev.
D} {\bf 68}, 116003.

\end{thebibliography}
\end{document}